\documentclass[10pt,conference,letterpaper]{IEEEtran}
\usepackage{times,epsfig}
\usepackage{amsmath,graphicx,algorithm,amssymb, amsthm, multirow, array, setspace}
\usepackage{float}
\usepackage{booktabs}

 \providecommand{\tabularnewline}{\\}

\title{Spatially Scalable Compressed Image Sensing with Hybrid Transform and Inter-layer Prediction Model}

\author{%
{Diego Valsesia, Enrico Magli }%
\vspace{1.6mm}\\
\fontsize{10}{10}\selectfont\itshape
Dipartimento di Elettronica e Telecomunicazioni\\
Politecnico di Torino, Italy\\
\fontsize{9}{9}\selectfont\ttfamily\upshape
diego.valsesia@polito.it\\
enrico.magli@polito.it%
\vspace{1.2mm}}

\begin{document}
\maketitle

\begin{figure}[b]
\parbox{\hsize}{\em
MMSP'13, Sept. 30 - Oct. 2, 2013, Pula (Sardinia), Italy.

978-1-4799-0125-8/13/\$31.00 \ \copyright 2013 IEEE.
}\end{figure}

\begin{abstract}
Compressive imaging is an emerging application of compressed sensing, devoted to acquisition, encoding and reconstruction of images using random projections as measurements. In this paper we propose a novel method to provide a scalable encoding of an image acquired by means of compressed sensing techniques. Two bit-streams are generated to provide two distinct quality levels: a low-resolution base layer and full-resolution enhancement layer. In the proposed method we exploit a fast preview of the image at the encoder in order to perform inter-layer prediction and encode the prediction residuals only. The proposed method successfully provides resolution and quality scalability with modest complexity and it provides gains in the quality of the reconstructed images with respect to separate encoding of the quality layers. Remarkably, we also show that the scheme can also provide significant gains with respect to a direct, non-scalable system, thus accomplishing two features at once: scalability and improved reconstruction performance.
\end{abstract}

\section{INTRODUCTION}
\label{sec:intro}
Compressed sensing (CS) has recently emerged as a technique to acquire signals having sparse or compressible representations by means of a small number of random projections. It is therefore a low complexity method to decrease the dimensionality of the signal of interest, and has spawned applications in fields as varied as sensor networks \cite{NowakSensor} and machine learning \cite{manifoldlearning}.
The use of compressed sensing for imaging has recently become a subject of great research interest. Compressive imaging is not motivated by reasons of compression efficiency, which is in fact lower than the one achievable by transform coding techniques such as JPEG or JPEG2000, whose rate-distortion performance is better. Instead, it is a promising acquisition framework, where one could think of directly acquiring random projections of an image by means of specialized hardware, such as spatial-multiplexing cameras (\emph{e.g.}, \cite{SPCamera}), with the great advantage of reducing the number of detectors needed. This is an appealing framework, e.g., for imaging beyond visible light, in which sensors can be much more expensive. As an example, an acquisition method for multidimensional signals (\emph{e.g.}, hyperspectral images) adopting CS is developed in \cite{Coluccia_jetcas}. 

Modern signal and image processing systems, however, not only acquire image data, but must communicate them to one or more receivers over a communication channel that may be prone to packet losses. Examples of such systems include wireless camera networks or the so-called smart camera platforms \cite{CITRIC} \cite{downesWisn}.  This, coupled with the potential heterogeneity of the receivers in terms of bandwidth, screen resolution and computational power, calls for the development of flexible signal representations that allow to successfully face these challenges.

In many multimedia applications, this is done by means of scalable coding. In particular, scalable coding is aimed at providing the ability to decode the signal in many different ways, so that the final user can choose the most suitable one, depending on factors such as the bit-rate that can be sustained by the transmission medium used, or the user's device (\emph{e.g.}, a battery-powered mobile, or a desktop computer, etc.).   
Typical approaches to scalable coding, such as the one adopted in the video standard H.264/SVC \cite{SVC}, aim at providing a scalable bit-stream in one or more of three ways: quality scalability, spatial resolution scalability, and temporal resolution scalability specifically for video signals.In this paper we focus on imaging, thus we will consider spatial resolution scalability and, briefly, quality scalability, although the main ideas carry over also to other multimedia contents such as audio and video with small modifications. 

Wireless camera networks comprising imaging nodes with limited computational capabilities can benefit from the low-complexity acquisition enabled by CS, whereas it would not be possible to perform image or video compression with classic methods. Keeping in mind the constraints dictated by the heterogeneity of receiver platforms, it is hence interesting to develop scalable coding techniques for systems adopting CS-based acquisition.
In applications other than CS, some approaches to scalable image coding benefit from having the image itself (or the wavelet transform like in EBCOT \cite{EBCOT}) available to encode it according to some paradigm of scalability. On the other hand, CS-based methods only have access to the random projections of the image. We bring this up in order to outline one of the main challenges in the development of a CS-based scalable signal representation: the necessity to avoid full recovery of the image at the encoder because of the typically large computational cost of this operation. Indeed, CS recovery algorithms based on $l_1$ norm minimization have a complexity that is cubic in the size of the signal to be reconstructed. One can resort to greedy algorithms such as OMP, but, still, the complexity is significant and it prevents from performing reconstruction immediately after the acquisition, as this would nullify the benefits of compressive imaging or of a CS-based low-complexity signal representation.

In this paper we show that it is possible to design a low-complexity scalable CS scheme. In particular, we propose an algorithm providing two or more layers at different spatial resolutions, without unduly increasing the complexity with respect to a conventional non-scalable CS scheme. We successfully exploit a predictive scheme that estimates the measurements of the enhancement layer from an interpolated preview of the image, that can be computed from the measurements of the base layer with a fast transform. This method allows to achieve higher performance, in terms of quality of the final reconstructed image, with respect to separate encoding of the two layers. Quite remarkably, not only the proposed scheme achieves very good scalability performance, but it also outperforms a single-layer non-scalable scheme at the same rate. Indeed, the proposed prediction-based approach employs a dual signal model, i.e. a sparsity model during the reconstruction stage, and a predictive smoothness model during the generation of the spatial scalability layers. This additional prior allows to consistently improve reconstruction quality, and opens the door to a new family of CS reconstruction schemes that make use of multiple signal models as in the present work.

 The problem of scalable encoding with CS has received little attention so far. Another approach to CS recovery is the progressive decoding outlined in \cite{BinnedCS}, where quantization with binning is exploited to improve the final quality. In that system, a preliminary recovery of the image from a first set of measurements is used to provide side information for the dequantization of the second set. Our approach is significantly different since it is aimed at providing two bit-streams carrying different quality levels. In fact, the preliminary recovery in \cite{BinnedCS} is not meant for visualisation as it produces a result with very poor quality. On the other hand, our approach offers a base layer with an acceptable quality level, and an enhancement layer to reach the best quality. Additionally, a low resolution version of the image can be recovered in a fast manner when a simple preview is needed.  

\section{BACKGROUND}
\label{sec:bkg}
CS is a novel theory for signal sensing and acquisition \cite{CS_donoho, candes2006compressive} allowing to acquire signals in an already compressed fashion, using fewer coefficients than dictated by the classical Nyquist-Shannon theory. Let us consider a signal $\mathbf{x} \in \mathbb{R}^n$, having a sparse representation under basis $\Psi \in \mathbb{R}^{n \times n}$:
$\mathbf{x} = \Psi\boldsymbol{\theta} \hspace{0.2cm} \mathrm{with} \hspace{0.2cm} \left\Vert \boldsymbol{\theta} \right\Vert_0 = k \ll n$, being $\left\Vert \boldsymbol{\theta} \right\Vert_0$ the $l_0$ norm of $\boldsymbol{\theta}$, \emph{i.e.}, the number of its nonzero entries.
We acquire measurements as a vector of random projections $\mathbf{y} = \Phi\mathbf{x} = \Phi\Psi\boldsymbol{\theta}$, $\mathbf{y} \in \mathbb{R}^m$, using a sensing matrix $\Phi \in \mathbb{R}^{m \times n}$.
The optimal way to recover the original signal from the measurements is to solve an optimization problem that minimises the $l_0$ norm of the signal in the domain where it is sparse. However, this problem is computationally intractable due to its NP-hard complexity, so it is common to consider a relaxed form using the $l_1$ norm, which can be solved by means of convex optimization techniques. In presence of bounded noise (\emph{e.g.}, due to quantization) it is common to consider an $l_2$ norm constraint using a bound $\epsilon$ on the noise norm, see \eqref{BPDN}.
\begin{align}
\label{BPDN}
\hat{\boldsymbol{\theta}} &= \arg \underset{\boldsymbol{\theta}}{\text{min}} \left\Vert \boldsymbol{\theta} \right\Vert_1 \hspace{.2cm} \text{subject to} \hspace{.2cm} \left\Vert \mathbf{y} - \Phi\Psi\boldsymbol{\theta} \right\Vert_2 \leq \epsilon 
\end{align}  
For the recovery of image signals, it is possible to use the previous method with a sparsifying basis like the wavelet transform. Another common approach is to use the Total Variation (TV) norm of the image. The TV norm of a 2D signal $x$, in its isotropic version, can be defined as 
\begin{align}
\label{TV}
TV(x) = \sum_{i,j} \sqrt{ \lvert x_{i+1,j} - x_{i,j} \rvert^2 + \lvert x_{i,j+1} - x_{i,j} \rvert^2}
\end{align}
Seeking to minimize the TV norm relies on the assumption that the gradient of the image is approximately sparse, hence the TV norm should be small.

The previous methods are successful provided that enough measurements have been acquired, typically $m = O\left( k\log \frac{n}{k} \right)$.

\section{SCALABLE CS ACQUISITION}
\label{sec: scalable_th}
\subsection{Encoder}

\begin{figure}[t]
  \centering
  \centerline{\includegraphics[width=0.95\linewidth]{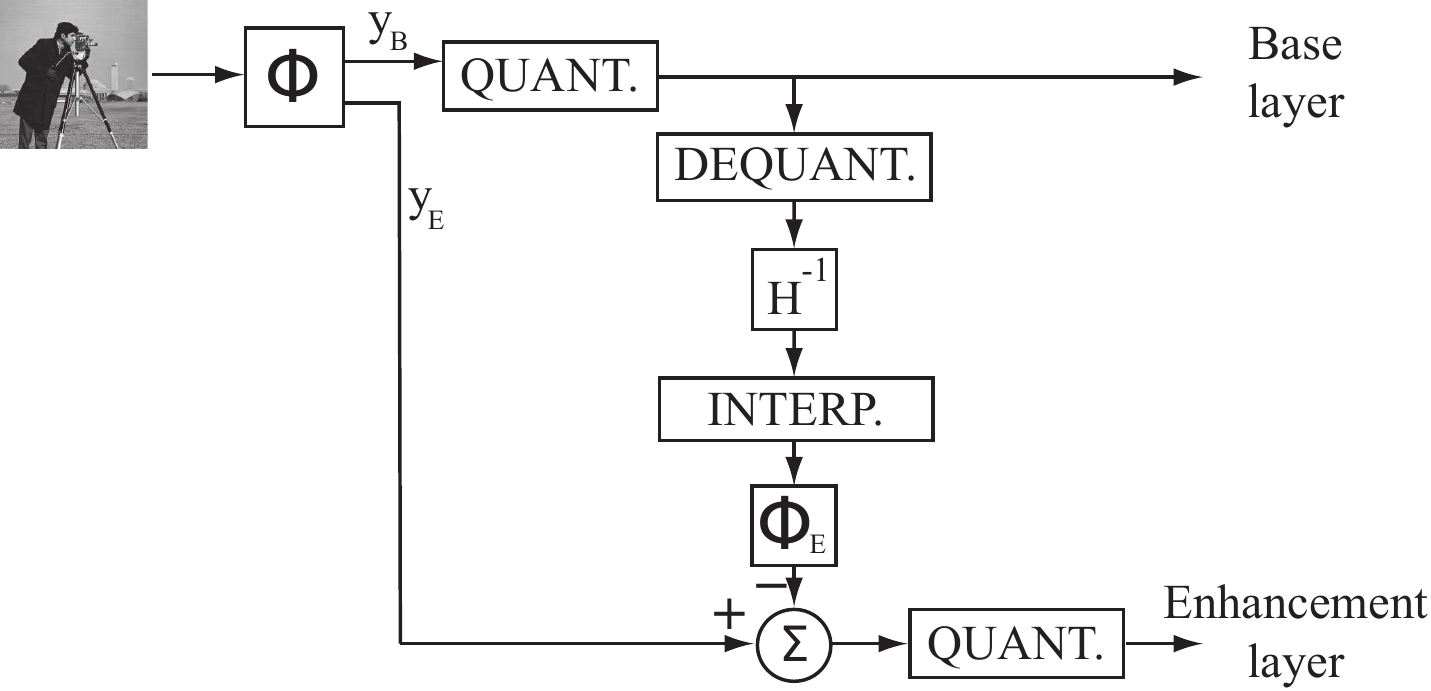}}
\medskip
\caption{\small{Scalable CS encoder.}}
\label{encoder}
\end{figure} 

In this section we outline the structure of the proposed scalable encoder providing two spatial scalability layers. A base layer provides a low resolution version of the original image that can be reconstructed from a limited number of measurements. The enhancement layer relies on the existing base layer to provide a high-resolution version of the image. Figure \ref{encoder} shows a block diagram representing the structure of the encoder.

The sensing mechanism makes use of two submatrices $\Phi_B$ and $\Phi_E$, with size $m_B \times n_B$ and $m_E \times n$ respectively, being $n$ the total number of pixels in the image and $n_B$ the number of pixels for the target resolution of the base layer. The submatrix used for the base layer has a smaller size in order to acquire measurements of a downsampled version of the original image. If we were to implement it in a practical system we would need to make it compatible with the size of $\Phi_E$, so we could think of building an augmented $\tilde{\Phi}_B$ to size $m_B \times n$ by introducing columns of zeros in the positions corresponding to the pixels that we don't want to acquire for the base layer. In this way we could build a single matrix $\Phi$ by stacking the augmented $\tilde{\Phi}_B$ and $\Phi_E$ upon each other.
The resulting matrix $\Phi$ overall contains only $\pm 1$ entries, so that it can be easily implemented on a hardware platform like the single-pixel camera of \cite{SPCamera}.
The measurements composing the base layer are obtained as $y_B=\tilde{\Phi}_B x$. The non-augmented version of $\Phi_B$ is the dual scale sensing matrix (DSS) proposed in \cite{CS-MUVI}. For our encoder, the key property of a DSS matrix is the ability to generate a fast preview of the image at a low resolution. A DSS matrix can be obtained as 
\begin{align}
\Phi_B = HD + F
\end{align}
where $H$ is a $m_B \times m_B$ Hadamard matrix (\emph{i.e.}, having $\pm 1$ entries only and such that $HH^T = m_B I$), $D$ an $n_B$-to-$m_B$ downsampling operator and $F$ contains a random pattern and it is such that $FU=FD^T=0$. This special construction guarantees that $\Phi_B$ is a good CS sensing matrix, in terms of RIP constant and recovery capability, thanks to the contribution of $F$. At the same time, it enables to generate a preview of the signal at low resolution because it minimizes the error between the downsampled version of the signal and the computed preview. The way to generate the preview is to compute $x_P = H^{-1}y_B$. Since $H$ is a Hadamard matrix, this is a fast operation that can be implemented by the fast Walsh-Hadamard transform algorithm. In our context, this enables the computation of a $\sqrt{m_B} \times \sqrt{m_B}$ preview image directly at the encoder.
The other submatrix $\Phi_E$ provides extra measurements $y_E = \Phi_E x$ allowing to generate the enhancement layer. However, $y_E$ is not transmitted directly. Instead, we use the preview of the image $x_P$ generated from the measurements of the base layer in order to predict the measurements of the enhancement layer, and then encode the prediction residuals. To this goal, we first interpolate the preview to bring it to the same resolution as the original image. In principle, any image interpolation method such as nearest neighbour, bilinear interpolation or even more sophisticated methods could be used. In a real implementation, one would choose the method according to the complexity they can afford for the encoder. The interpolated preview is then measured using $\Phi_E$ to obtain the prediction vector $y_{\mathrm{pred}}$. The difference $y_E - y_\mathrm{pred}$ is eventually encoded and transmitted. 

\begin{figure}[t]
  \centering
  \centerline{\includegraphics[width=0.95\linewidth]{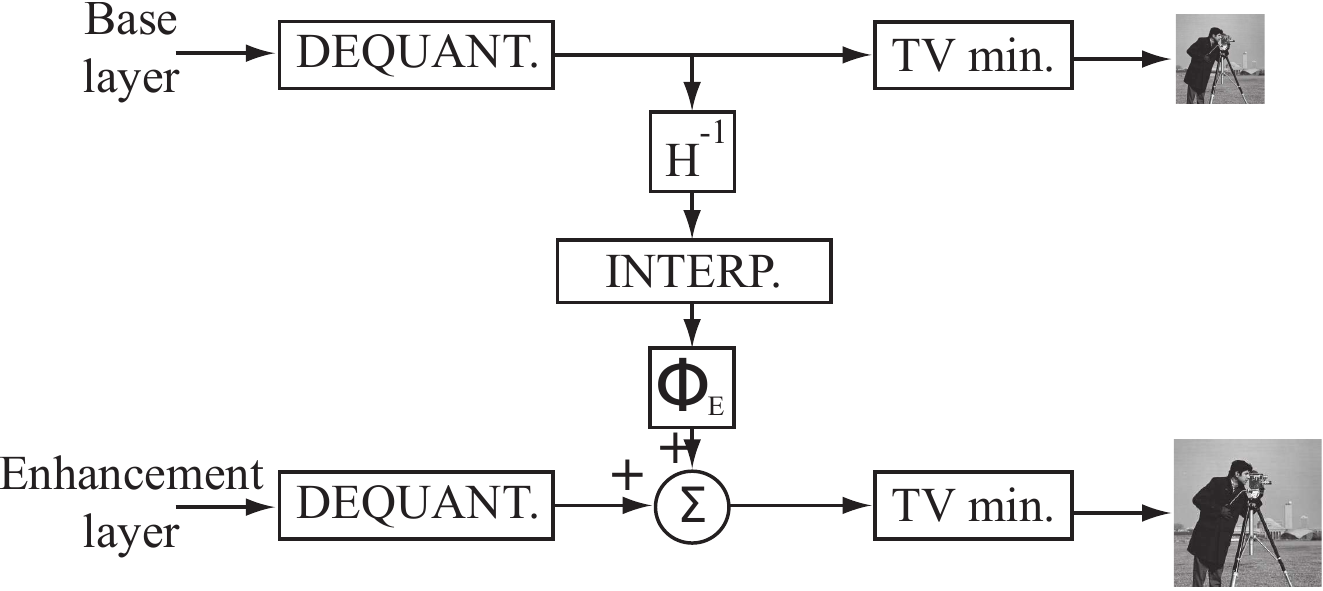}}
\medskip
\caption{\small{Scalable CS decoder.}}
\label{decoder}
\end{figure} 

\begin{figure}[t]
  \centering
  \centerline{\includegraphics[width=0.5\linewidth]{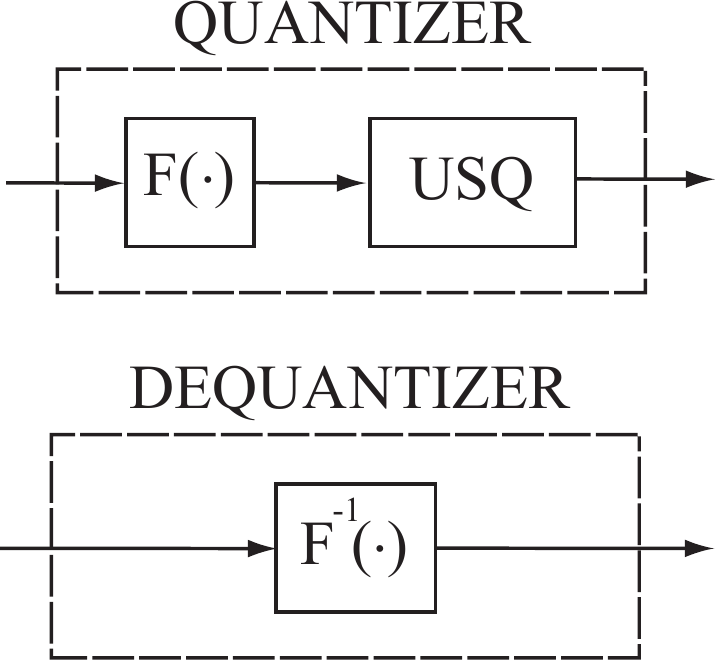}}
\medskip
\caption{\small{Companded quantizer. USQ is a uniform scalar quantizer over the $\left[0,1\right]$ interval.}}
\label{quantizer}
\end{figure}

The measurements of the base layer are quantized and dequatized before computing the preview in order to ensure alignment with respect to the decoder. Quantization can be performed with any technique, such as direct uniform scalar quantization of the measurements. However, we suggest a more efficient method that provides two benefits: (i) the quantizer is optimized to the distribution of the measurements, thus providing a lower quantization MSE, and (ii) the transmitted output forms an embedded bit-stream \cite{EquitzCover}, thus any truncation of it is optimal in the rate-distortion sense. This is performed by a companded quantizer, as shown in Fig. \ref{quantizer}, which consists of a compressor function, acting measurement-wise, to make the measurements approximately distributed according to a uniform probability density function, followed by a uniform scalar quantizer with $2^R$ reconstruction levels, being $R$ the quantization rate. We notice that the distribution of the measurements of the base layer is, with good approximation, Gaussian with zero mean and variance $\sigma_y^2$ due to the central limit theorem. Thus the compressor can be taken as the Gaussian cumulative density function:
\begin{align}
F(y_i) = \frac{1}{2}\left(1+\mathrm{erf}\left( \frac{y_i}{\sqrt{2\sigma_y^2}} \right) \right)
\end{align} 
Notice that since $F(y_i)$ has a uniform distribution, the uniform scalar quantizer is near-optimal. This quantizer produces an embedded code because of its tree structure. Truncating to $b$ bits the binary representation obtained from a $B$-bit uniform scalar quantizer yields the optimal $b$-bit representation (\emph{i.e.}, as if the quantizer was designed from the start for $b$ bits). This property provides quality scalability, in the sense that one could decide on-the-fly, during the transmission, to decrease the rate, without losing optimality in the quantization process.
Finally, the dequantization stage uses an expander function which is the inverse of the Gaussian cumulative density function.

\subsection{Decoder}
The decoder provides a reconstruction of the image with two resolution layers. The base layer works with a limited number of measurements $y_B$ to recover a low-resolution version of the image. This can be acceptable for a low-resolution user terminal or a user device that i subject to strict bandwidth constraints and thus cannot receive many measurements. On the other hand, the enhancement layer recovers the best possible quality by using the residual measurements after prediction. Notice that the decoder includes the same prediction branch as the encoder, and computes the measurements of an interpolated preview image from the measurements of the base layer. Recovery can be performed by using TV-norm minimization or $l_1$-norm minimization techniques, as desired. 

\section{EXPERIMENTAL PERFORMANCE}

\begin{figure}[t]
 \begin{minipage}[b]{.48\linewidth}
  \centering
  \centerline{\includegraphics[width=0.4\linewidth]{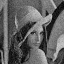}}
  \vspace{0.2cm}
  \centerline{(a) Fast preview (64$\times$64)}\medskip
\end{minipage}
\hfill
\begin{minipage}[b]{0.5\linewidth}
  \centering
  \centerline{\includegraphics[width=0.8\linewidth]{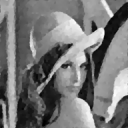}}
  \vspace{0.2cm}
  \centerline{(b) Base layer (128$\times$128)}\medskip
\end{minipage}
\\
\begin{minipage}[b]{\linewidth}
  \centering
  \vspace*{1cm}
  \centerline{\includegraphics[width=0.8\linewidth]{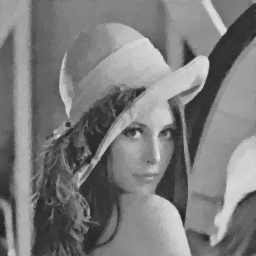}}
 \vspace{0.2cm}
  \centerline{(c) Enhancement layer (256$\times$256)}\medskip
\end{minipage}
\medskip
\vspace*{-0.5cm}
\caption{\small{Recovered images}}
\label{recovered}
\end{figure} 

In the previous section, we have outlined the structure of a scalable encoder adopting CS as an image acquisition method. We now evaluate its performance against a method that provides two quality layers by using separately $y_B$ and $y_E$, without any prediction. We also compare the proposed scheme against direct CS acquisition without any form of scalability. We shall discuss how the enhancement layer of the scalable scheme provides superior quality for the same bit-rate with respect to the direct approach. Hence, we emphasise how the scalable method accomplishes two features at once: improved reconstruction performance and two quality layers.

\begin{figure*}[t]
\vspace*{-0.5cm}
 \begin{minipage}[b]{.48\linewidth}
  \centering
  \centerline{\includegraphics[width=0.4\linewidth]{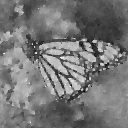}}
  \centerline{(a) Base layer }\medskip
\end{minipage}
\hfill
\begin{minipage}[b]{0.5\linewidth}
  \centering
  \centerline{\includegraphics[width=0.781\linewidth]{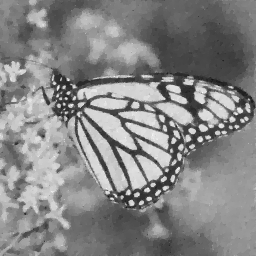}}
  \centerline{(b) Enhancement layer}\medskip
\end{minipage}
\medskip
\vspace*{-0.35cm}
\caption{\small{Base and Enhancement layer for Monarch}}
\label{monarch}
\end{figure*}

The tests were conducted on some standard test images like Lena, Cameraman, Goldhill, Boats, Monarch, Peppers, and the PSNR is used as quality metric. CS recovery is performed using TV minimization by means of the TVAL3 package \cite{TVAL3}. All images have full spatial resolution of $256 \times 256$ pixels. The scalable system computes a $64 \times 64$ preview, and the target resolutions for the base and enhancement layers are $128 \times 128$ and $256 \times 256$ respectively. The 64$\times$64 preview is interpolated to the full resolution before computing the predicted measurements by using a bilinear interpolator. Figure \ref{recovered} shows the actual recovered images for the preview, the base layer and the enhancement layer of image Lena (1.57 bpp overall, $0.3125$ bpp for the base layer), while figure \ref{monarch} shows the base and enhancement layers of Monarch at (1.68 bpp overall, $0.3125$ bpp for the base layer).

\begin{figure}[t]
  \centering
  \centerline{\includegraphics[width=0.99\linewidth]{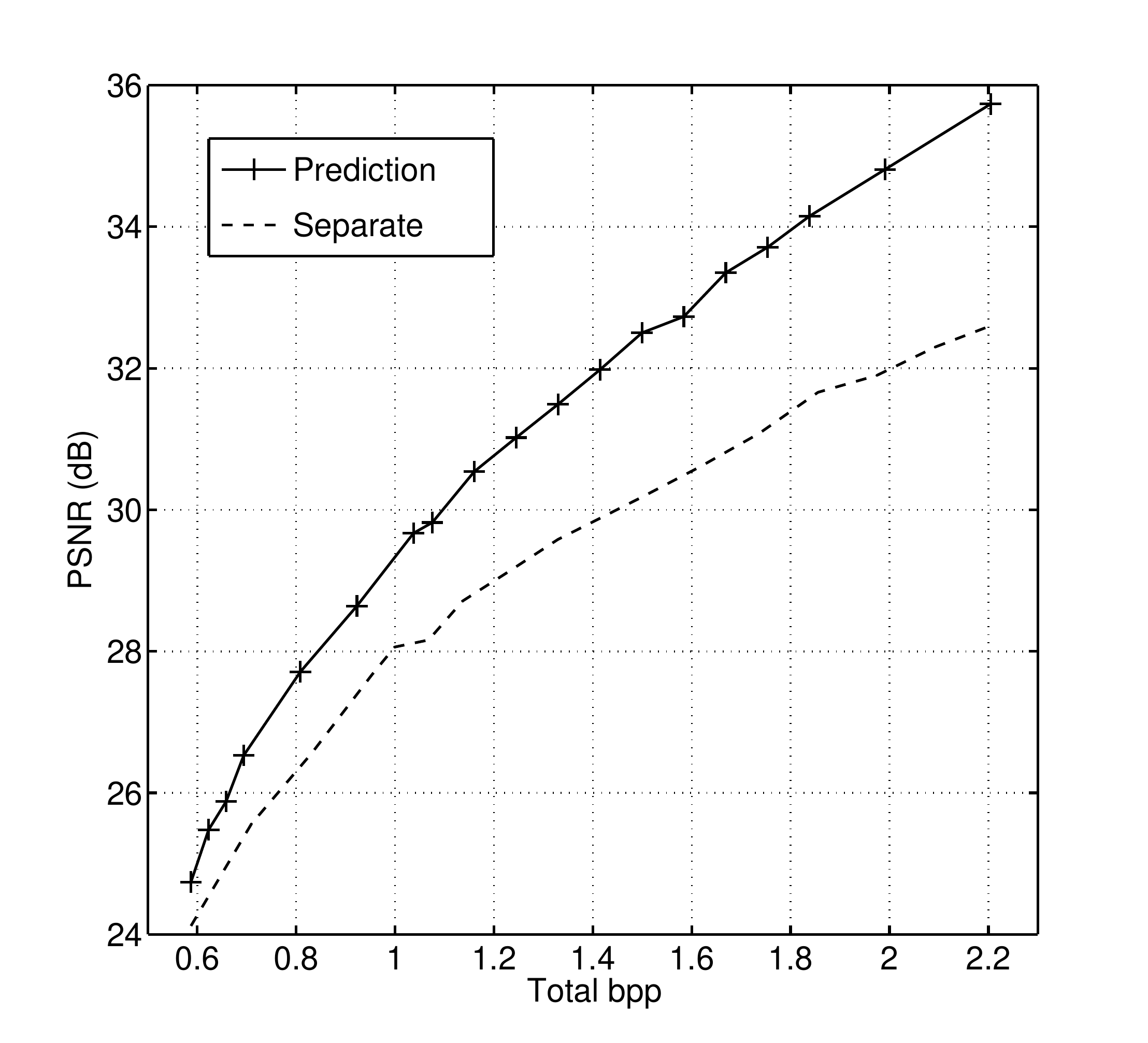}}
\medskip
\vspace*{-0.1cm}
\caption{\small{Predictive vs. separate encoding of the two layers. PSNR of the enhancement layer for the same total rate. Lena image of size $256 \times 256$.}}
\label{joint_vs_disjoint}
\end{figure}

Figure \ref{joint_vs_disjoint} reports the PSNR of the enhancement layer as a function of the total rate (base and enhancement layer) for the system using prediction from the preview image and for a system encoding the two layers separately. We can notice that the prediction mechanism is successful at providing significant quality gains. The plot has been obtained by choosing the best combination of number of measurements and quantization rate for each method, given a constraint on the total rate. The rate of the base layer has been fixed to $0.3125$ bpp, corresponding to $4096$ measurements quantized with 5 bits each.

\begin{table}[t]
\label{scal_table}
\vspace*{0.5cm}
\caption{PSNR non-scalable vs. enhancement layer of scalable. Scalable: Lena and Cameraman: $m_B=64^2$, $m_E =16500$, $R_B = 5$, $R_E = 5$, Goldhill, Boats, Peppers and Monarch: $m_B=64^2$, $m_E =18000$, $R_B = 5$, $R_E = 5$. Non-scalable: Lena: $m =11442$, $R = 9$, Cameraman: $m =14711$, $R = 7$, Goldhill, Boats, Peppers and Monarch: $m=13810$, $R = 8$. }
\vspace*{0.2cm}
\centerline{
\begin{tabular}{c c c c c}
\textbf{Image} & \textbf{Non-scalable} & \textbf{Scalable} & \textbf{Gain} & \textbf{bpp} \vspace*{0.1cm} \tabularnewline
\hline 
\textbf{Lena} & 30.48 dB & 32.86 dB & 2.38 dB & 1.57\tabularnewline
\hline 
\textbf{Cameraman} & 28.42 dB & 29.86 dB & 1.44 dB & 1.57\tabularnewline
\hline 
\textbf{Goldhill} & 26.69 dB & 27.76 dB & 1.07 dB & 1.68\tabularnewline
\hline 
\textbf{Boats} & 26.29 dB & 28.27 dB & 1.98 dB & 1.68\tabularnewline
\hline 
\textbf{Peppers} & 30.55 dB & 32.70 dB & 2.15 dB & 1.68\tabularnewline
\hline
\textbf{Monarch} & 27.46 dB & 29.52 dB & 2.06 dB & 1.68\tabularnewline
\hline
\end{tabular}}
\end{table}

\begin{figure}[t]
 \begin{minipage}[b]{.48\linewidth}
  \centering
  \centerline{\includegraphics[width=0.85\linewidth]{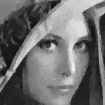}}
  \centerline{(a) Proposed method }\medskip
\end{minipage}
\hfill
\begin{minipage}[b]{0.5\linewidth}
  \centering
  \centerline{\includegraphics[width=0.84\linewidth]{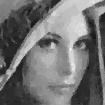}}
  \centerline{(b) Non-scalable method}\medskip
\end{minipage}
\medskip
\vspace*{-0.95cm}
\caption{\small{Detail of Lena at resolution 256$\times$256 and rate 1.57 bpp}}
\label{detail}
\end{figure}

As can be seen, a remarkable result of the proposed scheme is its ability to provide improved performance with respect to a non-scalable scheme that directly acquires a single layer of measurements, at the same total rate. In order to make this result more general, Table 1 compares the PSNR of the enhancement layer of the scalable method against the PSNR obtained with the direct, non-scalable method, at the same total rate. The table reports the results obtained using the best combination of number of measurements and quantization rate yielding the highest PSNR for a fixed rate for the non-scalable method, although in practice one would not know a priori the values of these two parameters. Both methods use the companded quantizer discussed in Sec. \ref{sec: scalable_th}. We notice that the scalable approach consistently provides gains. This can be also seen from Table 2 that reports the gain in PSNR at various rates for the test images. The main factor that enables the increase in performance is the use of a preview of the image to predict the measurements of the enhancement layer. Such preview allows to exploit additional prior information on the signal, which is smooth in its natural domain, thus allowing interpolation to do a good job at predicting the full-resolution version. This allows to exploit the correlation between the two sets of measurements of the base and enhancement layers, which is otherwise hidden in the compressed domain. We notice that our scheme could be considered as a more general technique to improve CS recovery performance, producing a scalable encoding of the signal as a side product. Indeed, one could think of ignoring the base layer and recover just the enhancement layer, solely to benefit of the performance gain. Figure \ref{detail} shows a detail of the Lena image recovered with a direct, non-scalable method and with the proposed scheme. The improvement in visual quality can be immediately assessed.

\begin{figure}[t]
  \centering
  \centerline{\includegraphics[width=0.85\linewidth]{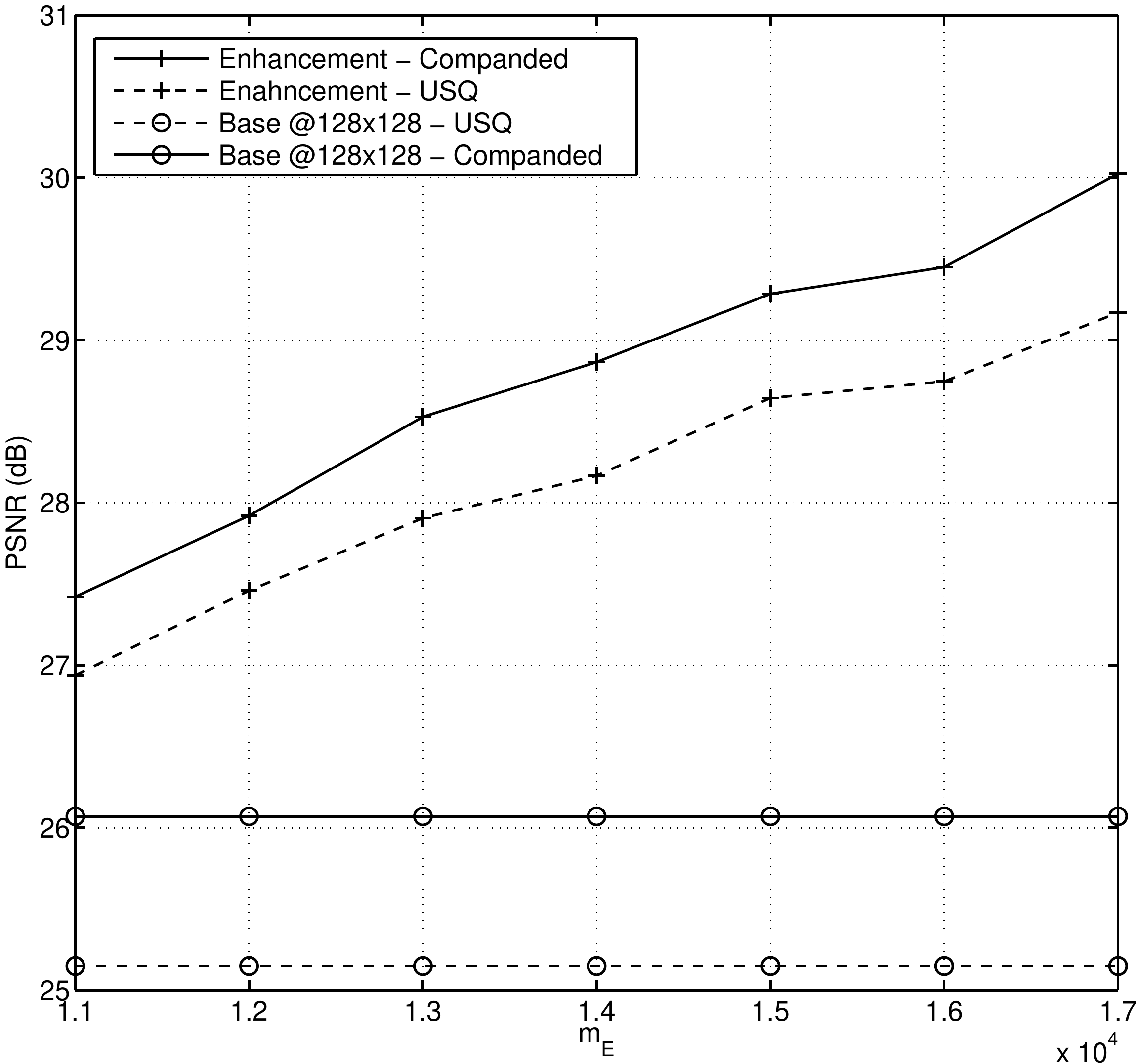}}
\medskip
\vspace*{-0.35cm}
\caption{\small{Companded quantizer vs. uniform scalar quantizer. Cameraman $256 \times 256$}}
\label{quantizer_gain}
\end{figure}

Finally, Fig. \ref{quantizer_gain} shows the gain achieved by the companded quantizer described in Sec. \ref{sec: scalable_th} with respect to a simple uniform scalar quantization of the measurements. Notice that direct uniform scalar quantization does not provide any optimization of the quantization process with respect to the distribution of the measurements, hence it cannot achieve the best rate-distortion performance and does not provide any sort of optimality if the bit-stream is truncated due to a variable-rate channel. Figure \ref{quantizer_gain} also reports the PSNR of the base layer.

\section{CONCLUSIONS}
In this paper we showed that it is possible to exploit the fast preview of an image from its compressive measurements, in order to provide a novel method for the scalable encoding of the original image. The role of the preview is to predict the measurements of the enhancement layer, allowing the encoding of the prediction residuals instead of the full measurements of the enhancement layer. We also proposed the use of a companded quantizer that leverages the Gaussian distribution of the compressive measurements in order to improve the quantization efficiency and, at the same time, provides an embedded bit-stream making quality scalability possible. Moreover, we showed that the proposed method provides gains in the quality of the reconstruction of the enhancement layer with respect to a direct non-scalable system at the same total rate. We conjecture that any signal having a smooth representation in the domain of acquisition, and with a frequency content that makes it possible to generate a downsampled preview without suffering from severe aliasing, can benefit from the proposed scheme in terms of improved reconstruction quality. The validation of this idea will be subject of future work.


\begin{table*}[t]
\caption{Gain in PSNR (dB) of the proposed method vs. a non-scalable reconstruction}
\centerline{
\begin{tabular}{c c c c c c c c c c c c c}
& \multicolumn{12}{c} {\textbf{Rates (bpp)}}
\tabularnewline
\cline{2-13}
 & \textbf{0.925}  & \textbf{1.000}  & \textbf{1.075}  & \textbf{1.150}  & \textbf{1.225}  & \textbf{1.300}  & \textbf{1.375}  & \textbf{1.450}  & \textbf{1.525}  & \textbf{1.600}  & \textbf{1.675}  & \textbf{1.750} \vspace*{0.1cm} \tabularnewline
\hline 
\textbf{Lena} & 0.94 & 1.27 & 1.33 & 1.86 & 1.88 & 2.22 & 2.36 & 2.23 & 2.32 & 2.49 & 2.63 & 2.80 \tabularnewline
\hline 
\textbf{Cameraman} & 0.07 & 0.23 & 0.21 & 0.24 & 0.65 & 0.55 & 1.15 & 1.11 &	1.24 & 1.51 & 1.54 & 1.92 \tabularnewline
\hline 
\textbf{Goldhill} & 0.43 & 0.45 & 0.63 & 0.67 & 0.78 & 0.85 & 1.04 & 1.07 & 1.06 & 1.01 & 1.38 & 1.24 \tabularnewline
\hline 
\textbf{Boats} & 0.32 & 0.72 & 0.81 & 0.97 & 1.04 & 1.32 & 1.45 & 1.34 & 1.68 & 1.57 & 1.83 & 2.15 \tabularnewline
\hline 
\textbf{Peppers} & 0.54 & 0.90 & 0.71 & 1.21 & 1.38 & 1.59 & 1.47 & 1.75 & 1.89 & 2.05 & 2.17 & 2.41 \tabularnewline
\hline 
\textbf{Monarch} & 0.38 & 0.79 & 1.37 &	1,38 & 1.25 & 1.74 & 1.56 & 1.89 & 1.99 & 2.03 & 2.11 &	2.39 \tabularnewline
\hline 
\end{tabular}}
\vspace*{0.5cm}
\end{table*}

\section*{Acknowledgment}
This work is supported by the European Research Council under the European Community's Seventh Framework Programme (FP7/2007-2013) / ERC Grant agreement n.279848.


\bibliographystyle{IEEEtran}
\bibliography{bibliography}

\end{document}